\documentclass[12pt,preprint]{aastex}

\slugcomment{Submitted to Ap. J. Letters}
\shorttitle{X-ray Spectrum of PSR B0656+14}
\shortauthors{Marshall \& Schulz}

\begin{document}

\title{Using the High Resolution X-ray Spectrum of PSR B0656+14
to Constrain the Chemical Composition of the Neutron Star Atmosphere}

\author{Herman L. Marshall and Norbert S. Schulz}
\affil{Center for Space Research, MIT, Cambridge, MA 02139}
\email{hermanm@space.mit.edu, nss@space.mit.edu}

\begin{abstract}

Observations of PSR B0656+14 using the Chandra Low Energy Transmission
Grating Spectrometer are presented.  The zeroth order events are pulsed
at an amplitude of 10 $\pm$ 2\% and the
image may be slightly extended.  The extended
emission is modelled as a Gaussian with a FWHM of about 0.75\arcsec,
for a linear size (at a distance of 760 pc) of 8.5 $\times 10^{15}$ cm.
In the absence of systematic errors in the detector point spread function,
the extended emission comprises $\la 50$\% of the observed flux in the
0.2-2.0 keV band, for a luminosity of $\la 3 \times 10^{32}$ erg s$^{-1}$.
The spectrum is well modelled by a dominant
blackbody with $T = 8.0 \pm 0.3 \times 10^5$ K and a
size of 22.5 $\pm$ 2.1 km in
addition to a harder component that is modelled as
a hotter and much smaller blackbody.
No significant absorption features are found in the spectrum that
might be expected from ionization edges of H or He
or bound-bound transitions of Fe in magnetized atmospheres.
Such features are expected to be deep but could
vary in position or strength with rotation phase.
There are no strong absorption features in the pulse-phased
spectra, however, so we conclude that the atmosphere is not dominated
by Fe or other heavy elements that would be partially ionized
at a temperature of $10^6$K.

\end{abstract}


\section{Introduction}

The X-ray spectra of many ``middle-aged'' and isolated
neutron stars are dominated by a thermal component,
so, besides temperatures and radii, they can in
principle be used to measure the atmospheric 
composition.  See \citet{bt97} for a summary 
of ROSAT observations of several neutron stars whose 
X-ray emission could be fitted to blackbody spectra.
The composition can, in turn, affect the estimated temperature
and size of the emission region
\citep{rr96,rutledge99} but
ROSAT spectra were insufficient to determine if there were
spectral features that might be expected if cyclotron lines
or heavy elements dominate the atmosphere.
For example, \cite{rrm97} showed that 
an iron-dominated atmosphere in a magnetic field
as strong as $10^{12}$ G would
show many narrow absorption features due to the distortion of
the energy levels.

PSR B0656+14 is an isolated radio pulsar with a period of about 384 ms,
found in the {\em Einstein} ultrasoft X-ray survey
by \cite{cordova}.
ROSAT observations showed weak pulsations
with an amplitude of $\sim$ 14\% \citep{finley92} and that the
X-ray spectrum appeared to be well fit with a simple black body
model with $T = 9 \times 10^5$ K.  From longer ROSAT
observations, \citet{possenti96} found that the spectrum appeared
slightly more complex and that the pulse shape changes with
energy.  \cite{possenti96} fitted a two component model
to the pulse light curve and X-ray spectrum, suggesting that
the hotter component is emission from the poles and the cooler
part comes from the equator but the hotter component could
instead be modelled by a steep power law
component with $\alpha =
3.5 \pm 0.4$ (where $f_{\nu} \propto \nu^{-\alpha}$).
\citet{edelstein} found a somewhat
lower temperature by combining {\em EUVE}, {\em ROSAT}, and
optical data.  \citet{greiveldinger} required two black body
components as well as a hard power law ($\alpha =
1.5 \pm 1.1$) to
describe the {\em ASCA} spectra extending to 5 keV.
The distance estimated from the pulsar's dispersion
measure is 760 pc \citep{tml93}.

We observed
PSR B0656+14 using the Chandra Low Energy Grating
Spectrometer (LETGS)
in order to detect narrow X-ray lines in absorption
that might be observed if the atmosphere is dominated by iron.
The energies of these absorption features should depend on
the average magnetic field strength, so we used the High
Resolution Camera Spectroscopy (HRC-S) detector which provide
event timing to better than 10 $\mu$s in order to obtain
pulse phased spectra.
We find no strong spectral lines and show that the LETGS
data are consistent with the previous results; the continuum
is well fit by a model involving two black body components.
Pulse phased spectra are used to
search for spectral features that might be phase dependent.

\section{Observations and Data Reduction}

\subsection{Imaging}
\label{sec:imaging}

PSR B0656+14 was observed with the LETGS on 28 November 1999
(JD - 2451000. = 510.99 - 511.43).  The level 0
were processed with CIAO version 2.0b.\footnote{CIAO
is the Chandra Interactive Analysis of Observations, a software
system developed by the Chandra X-ray Center.}
The exposure time was 38167 s.
The count rate in zeroth order was consistent with 0.19 count/s
for the entire observation.
The bandpass for the zeroth order image is hard to define due
to the low energy resolution of the HRC-S, so we estimated the
energy range over which one may integrate the spectrum-weighted
effective area to obtain 90\% of the observed count rate:
0.20 - 0.75 keV.  To obtain 99\% of the total count rate, the
bandpass should be extended to 0.16-1.0 keV.
The resultant image was
azimuthally symmetric, as expected for a point source.

We tested
the radial distribution of events against that of a comparable
LETG/HRC-S observation of a point source: Capella
(observation ID 1248).  Due to the dispersion by the fine and
coarse support structure, the point response function of the
LETGS zeroth order image depends slightly on the spectrum of
the source.  The zeroth order image yielded 180440 events
within 5\arcmin\ of the centroid in 85260 s.
The spectrum of Capella is soft but is
harder than that of the pulsar, so we divided the data
into two equal parts based on the pulse height distribution.
The average pulse height for the low half is quite similar
to that of PSR~B0656+14, but the radial profile differs
by no more than 1.5\% from the average profile using all events
so we used the average profile for comparison to the pulsar.
The encircled power was modelled as the sum of Gaussian and
exponential components, representing the Capella
data to better than 1.5\% over the 0-5\arcsec\ range:
\begin{equation}
\Phi(\theta) = (1-f) [ 1-e^{-\theta^2/\sigma_{PSF}^2/2} ] +
	f [1-e^{\theta/\theta_0}]  ,
\end{equation}

\noindent
where $\theta$ is the angle from the centroid, $f = 0.31$ is
the fraction of the power in the exponential component, $\sigma_{PSF}$
is the Gaussian width parameter of the point spread function (PSF),
and $\theta_0$ is the scaling length of the exponential component.
Capella is very bright, so
the background has a negligible contribution ($<$ 1\%) to the radial
profile out to 5\arcsec\ from the source centroid.
For the pulsar, background was estimated from an
annulus 8-10\arcsec\ from the source centroid.

\begin{figure}
\plotone{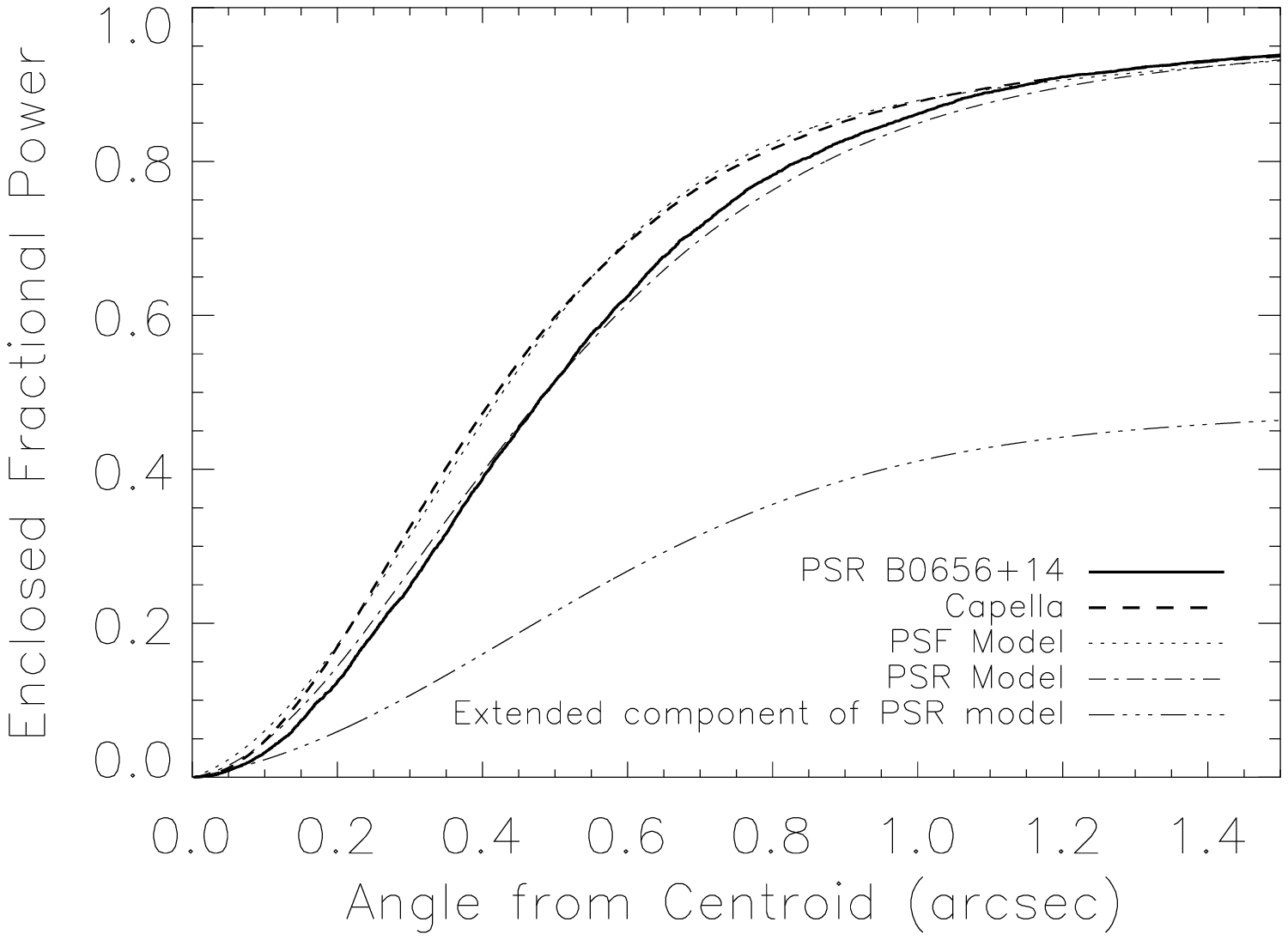}
\caption{Enclosed fractional power for the pulsar, PSR~B0656+14
(solid line), and for Capella (dashed line).
The pulsar's profile is somewhat broader than that of Capella,
giving an indication that there is some extended emission around
the neutron star.
A model of the point spread function (PSF) was fitted
to the Capella data.
The model of the pulsar's power profile consists of a PSF
and an extended emission modelled by a Gaussian with
$\sigma_{ext} = 0.32$\arcsec\ convolved with the PSF.
All profiles (except the extended component)
are normalized to unity at an angle of 5\arcsec\ from
the image centroid.
Given the possible systematic uncertainties in
early HRC-S data, we estimate that a point source contributes
$\ga$ 50\% of the source flux.
\label{fig:radialprofile} }
\end{figure}

The cumulative event radial profiles are given
in Fig.~\ref{fig:radialprofile}.
The profiles are normalized to unity at 5\arcsec.
The pulsar's radial distribution differs from Capella's by
$\sim$ 8.5\% at 0.5\arcsec, indicating that
the X-ray source may consist of a point source and some distributed
emission that is extended on a scale of order 0.5\arcsec.
A Smirnov test indicates that a 5\% difference between
the Capella and PSR~B0656+14
profiles would be significant at the $> 8\sigma$ level and
that differences of 1.6\% are significant at the 2$\sigma$ level.
We modelled the profile by combining a Gaussian model of the
extended emission with a PSF model.
The model of the extended emission was constructed
using the PSF model but increasing the Gaussian width
parameter by combining a second term in quadrature, $\sigma_{ext}$.
The model for the pulsar's radial distribution is matched to
about 2\% when $\sigma_{ext} = 0.32$\arcsec\ and the extended
component comprises 50\% of the total power.  The uncertainties
are difficult to characterize and are probably dominated
by systematic uncertainties in the PSF.  Given the
possible systematic uncertainties in these early HRC-S imaging data,
we estimate that an unresolved source accounts for $\ga$ 50\% of
the profile and that the FWHM of the extended emission is in the
range of 0.2-0.5\arcsec.

\subsection{Timing}

\begin{figure}
\plotone{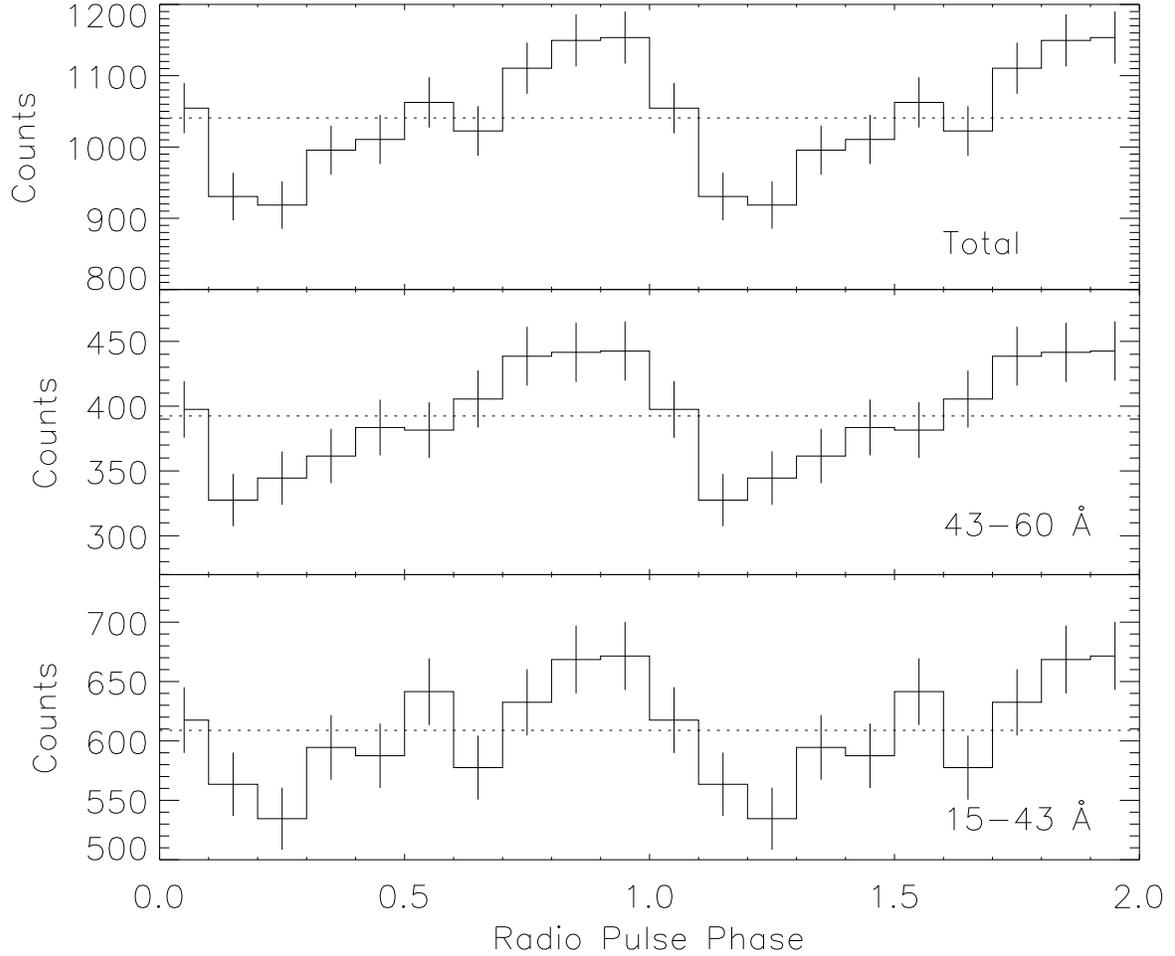}
\caption{The pulse profile of the zeroth order events.
{\em Bottom:} Hard bandpass (15-43 \AA, or 0.29-0.83 keV),
{\em middle:} soft bandpass (43-60 \AA\ or 0.21-0.29 keV),
{\em top:} total (15-60 \AA\ or 0.21-0.83 keV).
The data are replicated to the phase range 1-2 for presentation.
An unpulsed background level was subtracted from each profile.
The events are folded at the period given by an accurate
radio pulse ephemeris (Andrew Lyne, private communication):
384.89970 ms.  The pulse is asymmetric and the peak is centered
at phase 0.85 where zero phase is defined to be the peak of the
radio pulse.  Thus, it appears that the X-ray
pulse somewhat leads the radio pulse by about 0.15 in phase.
The soft and hard pulse profiles are quite similar.
\label{fig:pulse} }
\end{figure}

The zeroth order image was used to determine the phase and
period of the X-ray pulsations.  After applying timing corrections
to the solar system barycenter and correcting for the satellite's
position in orbit, we determined that the pulse period was
384.8990 ms with a 1 $\sigma$ uncertainty
of 0.0010 ms using a $\chi^2$ test on folded light curves.
This period is consistent with the predicted value, 384.89970 ms from
radio pulse monitoring (Andrew Lyne, 2001, priv. communication).
The pulse light curve shown in Fig.~\ref{fig:pulse} was folded
using the radio ephemeris.
The pulse appears to be asymmetric, as previously found by
\cite{possenti96}.  The modulation amplitude is 10 $\pm$ 2\%, defined
as $(R_{max} - R_{min}) / (R_{max} + R_{min})$, where $R_{max}$
and $R_{min}$ are the maximum and minimum of the pulse light
curve, respectively.  This value is consistent with the pulse
fractions
obtained by \cite{finley92} (14 $\pm$ 2\%) and
\cite{possenti96} (9 $\pm$ 1\%).

The radio phase is defined so that the peak of the radio pulse
occurs at a phase of zero.  Fig.~\ref{fig:pulse} shows that
the X-ray peak is at about phase 0.85, which is about
55\arcdeg\ out of phase with the radio phase.
The absolute timing HRC-S has been verified to an accuracy
of $<$ 0.001 s using the Crab pulsar
(Allyn Tennant, 2001, priv. comm.).

\subsection{Spectroscopic Data Reduction}
\label{sec:spectra}

The spectral data were reduced from standard event lists using
IDL using custom processing scripts; the method is quite similar to
standard processing using CIAO.  The procedure was to:
1) select data based on the criteria suggested by the Chandra X-ray Center
(CXC) LETGS calibration team\footnote{Chandra calibration information
is available at the CXC web site:{\tt http://cxc.harvard.edu/cal/}.
The HRC-S selection recommendation can be found at:
{\tt http://cxc.harvard.edu/cal/Links/Letg/\-User/Hrc\_bg/}.}
in order to reduce background,
2) determine the location of zeroth
order using one dimensional profiles fitted to Gaussians,
3) rotate events from sky coordinates to compensate for the telescope roll
and correct for a 0.54\arcsec\ offset between HRC-S plates 0 and 1,
4) compute
the dispersed grating coordinates ($m\lambda$ and $\theta$)
using the grating dispersion angles and the dispersion relation,
5) select ``source'' events
spatially within oppositely curving parabolas of the dispersion
line as suggested by the CXC LETG calibration team
\footnote{See {\tt http://cxc.harvard.edu/cal/Links/Letg/User/Hrc\_QE/EA/Wads}. },
6) eliminate data affected by detector gaps,
and 7) bin events at  $\Delta \lambda = $0.025\AA.

The effective area (EA) of the LETGS has undergone a few revisions
since launch both due to a recalculation of the LETG efficiencies
and due to a long series of in-flight calibration observations designed
to probe the HRC-S quantum efficiency.  We started with the
updated LETGS EA that was released by the CXC of 31 October 2000\footnote{
This effective area is available at
{\tt http://\-cxc.harvard.edu/\-cal/Links/Letg/User/Hrc\_QE/\-EA/\-correct\_ea/\-letgs\_NOGAP\_EA\_\-001031.mod}.}.
These LETG efficiencies were used, along with the transmission models
of the UV ion shield, to determine the effective areas for orders 2 through 5.
In a final step, an adjustment to the pre-flight calibration
was applied which was derived from an analysis of the spectrum of Mrk 478
\citep{mk478} and PKS 2155-304 (Marshall et al. 2002, in preparation).
The spectra of both targets are well fit by a simple power
law with Galactic absorption.
The adjustments affect primarily the spectrum at low energies
($E < 0.2$ keV), where high order contributions are beginning
to be important.

\begin{figure}
\epsscale{0.75}
\plotone{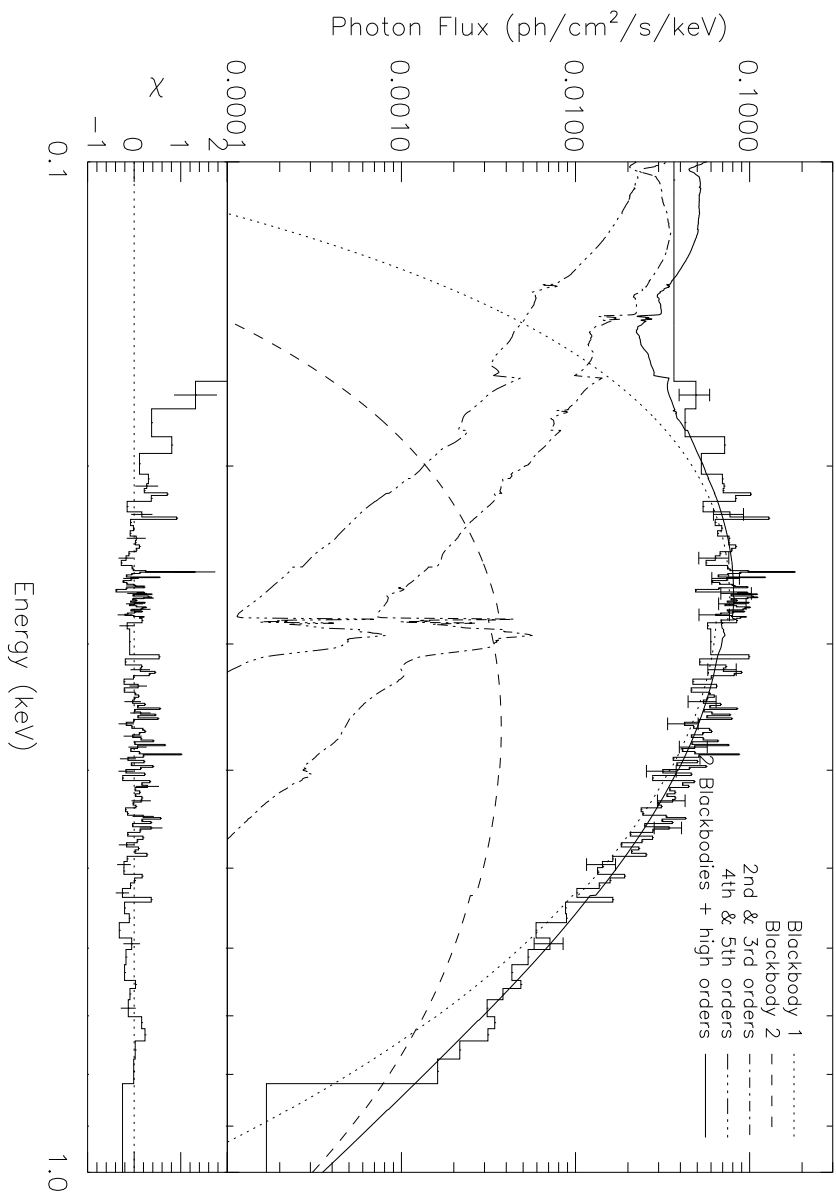}
\caption{The LETGS spectrum of PSR~B0656+14.  The bin sizes have
been varied to provide good signal in each energy bin; the
uncertainties are about 20\% everywhere.
{\em solid line:} a model consisting of two blackbody
components.  The lower panel shows the residuals as ratios
to the uncertainties.  High orders do not
contribute significantly for $E > 0.20$ keV while the data at
low energies ($E < 0.15$ keV) are best
modelled as the result of the sum of high orders.  The cooler
blackbody component dominates the total power while the hotter
component dominates the spectrum for $E > 0.8$ keV.
\label{fig:spectrum} }
\end{figure}

Directly totalling the observed fluxes over the 0.2-3.0 keV band
gives an observed flux of 1.01 $\pm 0.03 \times 10^{-11}$ erg
cm$^{-2}$ s$^{-1}$ for an absorbed luminosity of
7.0 $\pm 0.2 \times 10^{32}$ erg
s$^{-1}$.
The data were rebinned adaptively to provide a signal/noise
ratio of 5 in each
bin over the 0.10 to 2.0 keV range.  The spectrum, shown in
Fig.~\ref{fig:spectrum}, was estimated using
the first order EA only.  The contributions to the observed counts
due to high orders are estimated by folding a model for first order
through the high order EA and dividing by the first order EA.  This
procedure has the disadvantage that all the spectral features in the
high order EAs
are apparent but the advantage is that the result will match the
intrinsic spectrum well if first order dominates.  For this source,
high orders are negligible above 0.20 keV.

\section{Modelling the Spectrum}

\label{sec:models}

Following previous analyses, we modelled the continuum with two
blackbody components.  Our objective is primarily to define a
smooth continuum model to use as a baseline for line searches.
We did not include the hard power
law component found by \citet{greiveldinger} because its contribution
in the 0.2-1.0 keV band is negligible.
We exclude the data below 0.15 keV from
the fit where uncertainties in the high order
grating efficiencies can be important.
Gaussian statistics could be used for continuum modelling by rebinning
to obtain a signal/noise ratio of at least 5 in each bin, giving
reduced $\chi^2 = 1.13$,
acceptable at the 90\% level.  The distribution of the residuals
is not consistent with a Gaussian, however; the Kolmogorov test
rejects at the 99.99\% confidence level due to systematic
skewing of the residuals to positive values.  This may result
from systematic errors at the 5-10\% level
in the 0.45-0.55 keV portion of the spectrum.

\begin{figure}
\epsscale{1.0}
\plotone{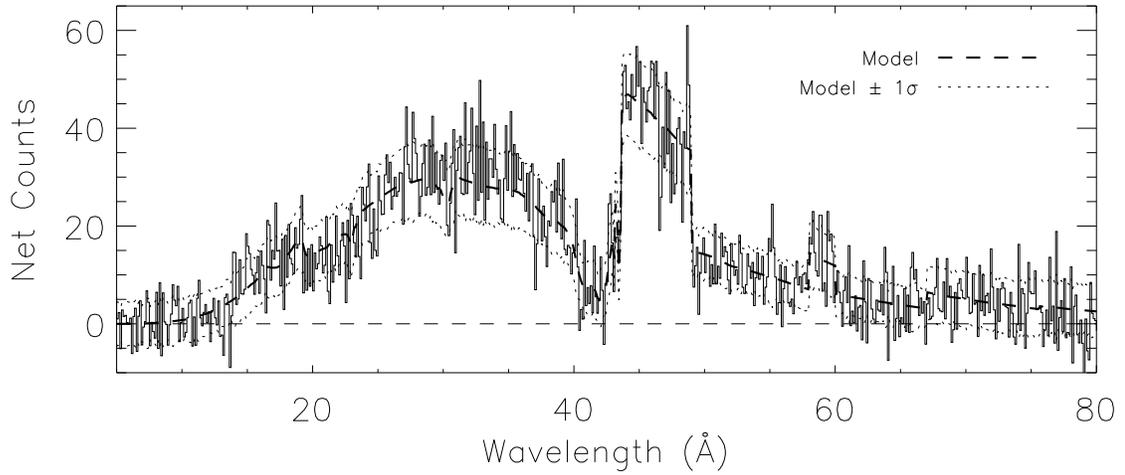}
\caption{The count spectrum of PSR~B0656+14 obtained with the
LETGS.  A binning of 0.125\AA\ was used to obtain sufficient
signal per bin to search for narrow features.
{\em Heavy dashed line:} expected count spectrum from
the model shown in figure~\ref{fig:spectrum}.
{\em Light dotted lines:} $\pm$ 1 $\sigma$ uncertainties about
the model.  The residuals are consistent with statistical
fluctuations about the model.  The sharp edges in the model
near 50 to 70\AA\ range are due to detector gaps.
\label{fig:countspec} }
\end{figure}

The best fit temperatures of the two blackbody components
were $8.0 \pm 0.3 \times 10^5$ and
$1.6 \pm 0.3 \times 10^6$ K and
the radii were 22.5 $\pm$ 2.1 and 1.7 $\pm$ 1.0 km, respectively.
The best-fit interstellar medium (ISM) column density, $N_H$,
was $1.73 \pm 0.18 \times 10^{20}$ cm$^{-2}$.
These parameters are close to those found by \citet{greiveldinger}.
The best fit model is plotted against the fitted data in
Fig.~\ref{fig:spectrum}.
The count spectrum (Fig.~\ref{fig:countspec}) was binned
at 0.125 \AA\ resolution in order to
search for narrow spectral features against the continuum model.
No significant absorption features were found.
A fit to a nonmagnetized H atmosphere \citep{zps96} gave
a much lower temperature, 230000 K, a very large radius,
413 $\pm$ 53 km and a much larger $N_H$, $2.6 \pm 0.2 \times
10^{20}$ cm$^{-2}$.

\begin{figure}
\epsscale{0.8}
\plotone{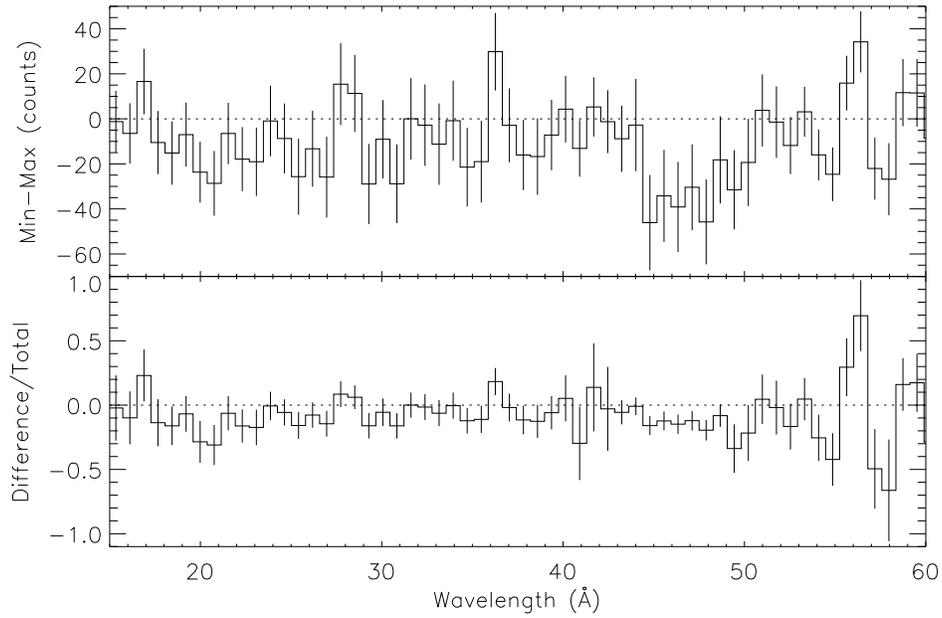}
\caption{Difference spectra in
0.775\AA\ bins.  The spectrum from the phase range 0.55-1.05
(see Fig.~\ref{fig:pulse})
is subtracted from the remaining phase range.
In the bottom panel, the difference spectrum is divided
by the model of the total spectrum, giving a fractional residual.
In the 45-50\AA\ band, there is a dip in the minimum spectrum
relative to that of pulse maximum, which is clearly detected
in the count spectrum.  The bottom panel shows that
the residuals are generally consistent with a
$\sim 8$\% difference between the two spectra that is independent
of wavelength.
\label{fig:specdiff} }
\end{figure}

The data were divided into pulse maximum and minimum
by phase: 0.05-0.55 and 0.55-1.05 (see Fig.~\ref{fig:pulse}).
The difference spectrum (Fig.~\ref{fig:specdiff})
shows a dip in the 45-50 \AA\ region.
The pulse amplitude spectrum was defined as the ratio
of the difference spectrum to the model of the pulse-averaged
spectrum and is also shown in Fig.~\ref{fig:specdiff}.  The result
is rather flat, consistent with being constant over this
wavelength band.

\section{Discussion and Summary}

Models of a neutron star atmosphere 
dominated by Fe in a strong magnetic field, such as
computed by \citet{rrm97}, should show deep narrow features
which we do not observe.  For a hydrogen atmosphere,
however, there are very few spectral features but
the radius of the fitted nonmagnetized H atmosphere is much too large.
In order to reduce the apparent radius to 13 km
(appropriate for a neutron star with
a mass of 1.4 $M_{\sun}$ and a radius of 10 km),
the distance would have to be absurdly small -- 24 pc --
given the large $N_H$ and the large distance from the dispersion measure.
For the blackbody model, however, the distance would have to be
about 440 pc, which is consistent with the estimate from
the pulsar's dispersion measure.
For a magnetized H atmosphere, results are similar to the
case with $B = 0$ but are not quite so extreme.
\citet{meyer94} found that the {\it ROSAT}
spectrum of Geminga fit a blackbody with $T = 7.6 \times 10^5$ K
while a magnetized H atmosphere gave $T = 5.0 \times 10^5$ K for
$B = 4.7 \times 10^{12}$ G.  The magnetic field estimated for PSR~B0656+14
is the same \citep{tml93}, so one might expect a similar
reduction when fitting the PSR B0656+14 spectrum.  In order
to generate the same observed flux, the radius of
the emission region would have to increase by a factor of 2.3 to
52 $\pm$ 5 km so the distance would have to be reduced to 190 pc.
A distance to less than 200 pc can be definitely
ruled out using the dispersion measure and a more refined model
of the local ISM (J. Cordes, 2002, private communication).

We find no significant absorption features in the
pulse-averaged or pulse-phased spectra over the 0.15-0.80 keV band.
Thus, we can rule out
electron and proton cyclotron resonance lines, which
should be rather deep and possibly broadened,
according to models by \cite{zane01} and \cite{holai01}.
The proton cyclotron line would be expected at an observed
energy $E_o = E_p / (1+z) = \hbar eB/(m_p c)/(1+z) =  0.0063 /
(1+z) B_{12}$ keV, where $B_{12} = B / (10^{12}$ G) and
$z = (1-2GM/Rc^2)^{-1/2} - 1$ is the surface redshift.
Substituting the electron mass for the proton mass, $m_p$,
gives the electron cyclotron line energy
$E_e / (1+z) = \hbar eB/(m_e c)/(1+z) =  1.16 /
(1+z) B_{12}$ keV.
Thus,
we eliminate the ranges $0.017 < B_{12} < 0.090$ and $30 < B_{12} < 165$
if $z = 0.30$ (for a neutron star of
1.4 M$_{\sun}$ and a radius of 10 km).
These ranges do not exclude the magnetic field estimate
from the dipole braking model: $B_{12} = 4.7$
\citep{tml93}.  The lower limits can be reduced if the bulk of the
emission comes from the equatorial zone where the magnetic field is
$\sim 50$\% of the polar value or if the absorbing plasma is far
off of the neutron surface.

For this dipole field estimate, a
light element atmosphere should show H or He ionization edges
in the 0.15-0.80 keV bandpass.  \cite{ls97} give an estimate for
the ionization edge energy and its dependence on the
atomic number $Z$:
\begin{equation}
E_Z = 4.4 Z^2 [ \ln ( \frac{B}{Z^2 B_0} ) ]^2 {\rm ~~eV}
\end{equation}

\noindent
where $B_0 = 2.35 \times 10^{9}$ G.
A He {\sc i} edge would be expected in the 0.5-0.7 keV range.
We do not detect this edge, which may not be surprising, as helium
may have settled in the atmosphere to high optical depths.
For surface temperatures of 9-22 $\times 10^5$ K (allowing
for variation between the pole and the equator), He is almost
completely neutral while H is 7-22\% ionized.
So, we expect to find the H {\sc i} edge between 0.20 and 0.25 keV
(depending on the surface gravity) but it is not observed.
The spectrum is dominated by the
cooler thermal component, however, which is presumably emission from the
equatorial zone in which $B$ is $\sim 2\times$ weaker than at the pole.
Thus, the edge may be found in the 0.16-0.21 keV region, where
the signal/noise ratio is poor and
high order contributions to the LETGS spectrum become important.

Most of a pulsar's spin-down luminosity is thought to be carried  off
as a relativistic electron/positron wind.  When this wind is confined
by external pressure, a shock forms and the relativistic particles
radiate synchrotron emission in the form of a pulsar
wind nebula (PWN). The Crab Nebula
(Hester et al. 1995, Weisskopf et al. 2000) is the best known example.
PWNe have been observed around at least 6 pulsars so far \citep{bt97,lu02}.
Claims from ASCA observations of vastly extended PWNe ($>$ 10-20\arcmin)
in this and other pulsars
with luminosities of the order of $10^{32}$ erg s$^{-1}$
have not been confirmed 
by {\em ROSAT} and {\em BeppoSAX} \citep{becker99}.
We also do not detect such a large extended nebula around PSR B0656+14.
The image appears slightly extended, consistent with a point
source comprising $\ga 50$\% of the total 0.1-3 keV luminosity.
The remainder, with an unabsorbed luminosity of
$\la 3 \times 10^{32}$ erg s$^{-1}$
can be modelled with a Gaussian with a FWHM of 0.75\arcsec\
corresponding to a size of 8.5 $\times 10^{15}$ cm.
This extended emission is consistent with the result of \citet{becker99},
who found that a PWN around PSR B0656+14 must have an
extent of less than 10\arcsec.  Based on its spin-down age, $\sim 10^5$ yr,
PSR B0656+14 is an order of magnitude older
than the Vela pulsar, the oldest pulsar around which a PWN has
been detected \citep{bt97}. If we assume a similar emission efficiency
for PSR B0656+14 as for the Vela pular ($\sim 0.04\%$),
we can set an upper limit to the expected luminosity of the PWN of
1.5$\times 10^{33}$ erg s$^{-1}$, consistent
with the observed value.

\acknowledgments

We are extremely grateful to Prof. Claude R. Canizares,
the Principal Investigator of the {\em Chandra} High
Energy Transmission Grating Spectrometer, for
allocating part of his guaranteed time to this observation.
We thank the referee for comments that have
resulted in significant improvements to the paper.
This work has been supported in part under NASA contract
SAO SV1-61010.

\clearpage

\end{document}